\documentclass{article}
\usepackage{spconf,amsmath,graphicx}
\usepackage{array}
\usepackage{float}
\usepackage{multirow}
\usepackage{amssymb}

\title{Analysis of Multilingual Sequence-to-Sequence \\ speech recognition systems}
\name{\parbox{1.0\linewidth}{\center Martin Karafi\'{a}t$^1$, Murali Karthick Baskar$^1$, Shinji Watanabe$^2$,
  Takaaki Hori$^3$ \newline Matthew Wiesner$^2$ and Jan ``Honza''
  \v{C}ernock\'{y}$^1$\thanks{The work reported here was carried out during the 2018 Jelinek Memorial Summer Workshop on Speech and Language Technologies, supported by Johns Hopkins University via gifts from Microsoft, Amazon, Google, Facebook, and MERL/Mitsubishi Electric. It was also supported by  Czech Ministry of Education, Youth and Sports from the National Programme of Sustainability (NPU II) project "IT4Innovations excellence in science - LQ1602" and by the Office of the Director of National Intelligence (ODNI), Intelligence Advanced Research Projects Activity (IARPA) MATERIAL program, via Air Force Research Laboratory (AFRL) contract \# FA8650-17-C-9118. The views and conclusions contained herein are those of the authors and should not be interpreted as necessarily representing the official policies, either expressed or implied, of ODNI, IARPA, AFRL or the U.S. Government. 
}}}
\address{Brno University of Technology$^1$, John Hopkins
  University$^2$ \\
  Mitsubishi Electric Research Laboratories (MERL)$^3$}
\begin{document}
\ninept
\maketitle
\begin{abstract}
This paper investigates the applications of various  multilingual approaches developed in conventional hidden Markov model (HMM) systems to sequence-to-sequence (seq2seq) automatic speech recognition (ASR).
On a set composed of Babel data, we first show the effectiveness of multi-lingual training with stacked bottle-neck (SBN) features.
Then we explore various architectures and training strategies of multi-lingual seq2seq models based on CTC-attention networks including combinations of output layer, CTC and/or attention component re-training. 
We also investigate the effectiveness of language-transfer learning in a very low resource scenario when the target language is not included in the original multi-lingual training data. 
Interestingly, we found multilingual features superior to multilingual models, and this finding suggests that we can efficiently combine the benefits of the HMM system with the seq2seq system through these multilingual feature techniques.
\end{abstract}
\section{Introduction}
\label{sec:intro}

The sequence-to-sequence (seq2seq) model proposed in
\cite{sutskever2014sequence,bahdanau2014neural,cho2014learning} 
is a neural network (NN) architecture for performing sequence classification.  Later, it was also adopted to perform speech recognition 
~\cite{chorowski2015attention,graves2014towards,graves2012supervised}. 
The model allows to integrate the main blocks of ASR (acoustic model, alignment model and language model) into a single neural network architecture.
The recent ASR advancements in connectionist temporal classification
(CTC) \cite{graves2012supervised,graves2014towards} and attention
\cite{chorowski2015attention,chan2016listen} based approaches have generated significant 
interest in speech community to use seq2seq models. However, outperforming conventional hybrid
RNN/DNN-HMM models with seq2seq requires a huge amount of data \cite{rosenberg2017end}.
Intuitively, this is due to the range of roles this model needs to perform: alignment
and language modeling along with acoustic to character label mapping. 

Multilingual approaches have been used in hybrid RNN/DNN-HMM
systems  for tackling the problem of low-resource data. 
These include language adaptive training and shared
layer retraining \cite{tong2017investigation}. Parameter sharing investigated in our previous work  \cite{karafiat2016multilingual} seems to be the most beneficial.

Existing multilingual approaches for seq2seq modeling mainly focus on {\em CTC}. 
A multilingual CTC  proposed in \cite{tong2017multilingual} uses a universal phone set, FST decoder and language model. 
The authors also use a linear hidden unit contribution (LHUC) \cite{swietojanski2014learning} technique to rescale the hidden unit outputs for each language as
a way to adapt to a particular language. 
Another work \cite{muller2017language} on multilingual CTC shows the importance of language adaptive vectors as auxiliary input to the encoder in multilingual CTC model. The decoder
used here is based on simple greedy search of applying $argmax$ on every time frame. An extensive analysis of
multilingual CTC performance with limited data  is performed in \cite{dalmia:ICASSP:2018}. 
Here, the authors use a word level FST decoder integrated with CTC during decoding.

On a similar front, {\em attention models} are explored within a multilingual
setup in \cite{watanabe2017language,toshniwal:ICASSP:2018:multilingualE2E}, where an attempt was made to build an attention-based seq2seq model from multiple languages. The data is just pulled together assuming the target languages are seen during the training. 
Although our prior study \cite{cho:SLT:2018} performs a preliminary investigation of transfer learning techniques to address the unseen languages during training, this is not an intensive study of covering various multi-lingual techniques.

In this paper, we explore the multilingual training
approaches~\cite{tuske2014multilingual,karafiat:SLT:2016:MultBLSTM} 
in hybrid RNN/DNN-HMMs and we  incorporate them into the seq2seq
models. In our recent work~\cite{karafiat:SLT:2016:MultBLSTM}, we showed the
multilingual acoustic models (BLSTM particularly) to be superior to
multilingual acoustic features in RNN/DNN-HMM systems. Consequently, similar experiments are
performed in this paper on a sequence-to-sequence scheme.
 
The main motivation and contribution behind this work is as follows:
\begin{itemize}
\item To incorporate the existing multilingual approaches in a joint CTC-attention~\cite{watanabe2017hybrid} framework.
\item To compare various multilingual approaches:   multilingual features, model architectures, and transfer learning.
\end {itemize}

\section{Sequence-to-Sequence Model}\label{sec:seq2seq}
In this work, we use the attention based approach \cite{bahdanau2014neural}
as it provides an effective methodology to perform sequence-to-sequence
training. Considering the limitations of attention in performing
monotonic alignment \cite{sperber18interspeech,DBLP:journals/corr/abs-1712-05382},
we choose to use CTC loss function to aid the attention mechanism in both training and decoding.

Let $X = (\mathbf{x}_t | t=1, \dots, T)$ be a $T$-length speech feature sequence and $C= (c_l | l=1, \dots, L)$ be an $L$-length grapheme sequence.
A multi-objective learning framework $\mathcal{L}_{\text{mol}}$ proposed in \cite{watanabe2017hybrid} is used in this work to unify 
attention loss $p_{\text{att}}(C|X)$ and CTC loss $p_{\text{ctc}}(C|X)$ with a linear interpolation weight $\lambda$, as follows:
\begin{equation}
\label{eq:mtl}
\mathcal{L}_{\text{mod}}=\lambda\,\log\,p_{\text{ctc}}(C|X)+(1-\lambda)\,\log\,p_{\text{att}}^{*}(C|X).
\end{equation}
The unified model benefits from both effective sequence level training and the monotonic afforded by the CTC loss. 

$p_{\text{att}}\left(C|X\right)$ represents the posterior probability of character label sequence $C$ w.r.t input sequence
$X$ based on the attention approach, which is decomposed with the probabilistic chain rule, as follows:
\begin{equation}
\label{eq:att}
p_{\text{att}}^{*}\left(C|X\right)\thickapprox \prod _{l=1} ^L p\left(c_{l}|c_{1}^{*},....,c_{l-1}^{*},\ X\right),
\end{equation}
where $c_{l}^{*}$ denotes the ground truth history.
Detailed explanation of the attention mechanism is given later.

Similarly, $p_{\text{ctc}}\left(C|X\right)$ represents the CTC posterior probability: 
\begin{equation}
p_{\text{ctc}}\left(C|X\right)\thickapprox \sum _{Z \in \mathcal{Z}(C)} p(Z|X),
\end{equation}
where $Z= (\mathbf{z}_t | t=1, \dots, T)$ is a CTC state sequence composed of the original grapheme set and the additional blank symbol. 
$\mathcal{Z}(C)$ is a set of all possible sequences given the character sequence $C$.
%

%

The following paragraphs explain the encoder, attention decoder, CTC, and joint decoding used in our approach.

\subsubsection*{Encoder} 
In our approach, both CTC and attention use the same encoder function: 
\begin{equation}
 \label{eq:enc}
 \mathbf{h}_t = \text{Encoder}(X),
\end{equation}
where $\mathbf{h}_t $ is an encoder output state at $t$.
As $\text{Encoder}(\cdot)$, we use bidirectional LSTM (BLSTM). 


\subsubsection*{Attention Decoder:}
Location-aware attention mechanism \cite{locatt} is used in this work.
The output of location-aware attention is:
\begin{equation}
\label{locatt}
a_{lt}=\text{LocationAttention}\left(\left\{ a_{l-1, t'}\right\} _{t'=1}^{T}, \mathbf{q}_{l-1}, \mathbf{h}_{t}\right).
\end{equation}
Here,  $a_{lt}$ acts as  attention weight, $\mathbf{q}_{l-1}$ denotes the decoder hidden state, and $\mathbf{h}_{t}$ is the encoder output state defined in~\eqref{eq:enc}.
The location-attention function is given by a convolution and  maps the attention weight of the previous label $a_{l-1}$ to a multi channel view $\mathbf{f}_{t}$ for better representation: 
\begin{align}
\mathbf{f}_{t} & =\mathbf{K}*\text{\ensuremath{\mathbf{a}}}_{l-1}, \label{conv_feat} \\
e_{lt} & = \mathbf{g}^{T} \tanh(\text{Lin}(\mathbf{q}_{l-1})+ \text{Lin}(\mathbf{h}_{t})+\text{LinB}(\mathbf{f}_{t})), \label{energy}\\
a_{lt} & = \text{Softmax}(\{e_{lt}\}_{t=1}^{T}) \label{softmax}
\end{align}
Here, \eqref{energy} provides the unnormalized attention vectors computed with the learnable vector $\mathbf{g}$, linear transformation $\text{Lin}(\cdot)$, and affine transformation $\text{LinB}(\cdot)$.
Normalized attention weight are obtained in \eqref{softmax} by a standard $\text{Softmax}(\cdot)$ operation.   
Finally, the context vector $\mathbf{r}_{l}$ is obtained as a weighted sum  of the encoder output states $\mathbf{h}_{t}$ over all  frames, with the attention weight:
\begin{equation}
\mathbf{r}_{l}=\sum_{t=1}^{T}a_{lt}\mathbf{h}_{t}.
\end{equation}
The decoder function is an LSTM layer which decodes the next character output label $c_{l}$ from their previous label $c_{l-1}$, hidden state of the decoder ${\bf q}_{l-1}$ and attention output $\mathbf{r}_{l}$:
\begin{equation}
p\left(c_{l}|c_{1},....,c_{l-1},\ X\right)=\text{Decoder}(\mathbf{r}_{l},q_{l-1},c_{l-1})
\end{equation}
This equation is incrementally applied to form $p_{\text{att}}^{*}$ in~\eqref{eq:att}.

\subsubsection*{Connectionist temporal classification (CTC):}
Unlike the attention approach, CTC does not use any specific decoder network. 
Instead, it invokes two important components to perform character level training and decoding: the first one is an RNN-based encoding module $p(Z|X)$.
The second component contains a language model and state transition module. 
The CTC formalism is a special case \cite{graves2012supervised} of hybrid DNN-HMM framework with the Bayes rule applied to obtain $p(C|X)$.

\subsubsection*{Joint decoding:}
Once we have both CTC and attention-based seq2seq models trained, both are jointly used for decoding as below:
\begin{align}
\begin{split}\log p_{\text{hyp}}(c_{l}|c_{1},....,c_{l-1}, X)=\\
\alpha\log p_{\text{ctc}}(c_{l}|c_{1},....,c_{l-1}, X)\\
+(1-\alpha)\log p_{\text{att}}(c_{l}|c_{1},....,c_{l-1}, X)
\end{split}
\label{joint decoding}
\end{align}
Here $\log p_{\text{hyp}}$ is a final score used during beam search.
$\alpha$ controls the weight between attention and CTC models.
$\alpha$ and multi-task learning weight $\lambda$ in~\eqref{eq:mtl} are set differently in our experiments.

\section{Data}
\label{sec:data}

The experiments are conducted using the BABEL speech corpus collected during  the IARPA Babel program. The corpus is mainly composed of conversational telephone speech (CTS) but some scripted recordings and far field recordings are present as well. 
Table~\ref{tab:data} presents the details of the languages used  for training and evaluation in this work.
We decided to evaluate also on training languages to see effect of multilingual training on training
languages. Therefore, Tok Pisin, Georgian from ``train'' languages and
Assamese, Swahili from ``target'' languages are taken
for evaluation.

\begin{table}[tb]
\scalebox{1.15}{%
\centering{}{\tiny{}}%
\begin{tabular}{|c|c|c|c|c|c|c|}
\hline 
\multirow{2}{*}{{\tiny{}Usage}} & \multirow{2}{*}{{\tiny{}Language}} & \multicolumn{2}{c|}{{\tiny{}Train}} & \multicolumn{2}{c|}{{\tiny{}Eval}} & \multirow{2}{*}{{\tiny{}\# of }}\tabularnewline
\cline{3-6} 
 &  & {\tiny{}\# spkrs.} & {\tiny{}\# hours} & {\tiny{}\# spkrs.} & {\tiny{}\# hours} & {\tiny{}characters }\tabularnewline
\hline 
\hline 
\multirow{10}{*}{{\tiny{}Train}} & {\tiny{}Cantonese } & {\tiny{}952} & {\tiny{}126.73} & {\tiny{}120} & {\tiny{}17.71} & {\tiny{}3302}\tabularnewline
\cline{2-7} 
 & {\tiny{}Bengali } & {\tiny{}720} & {\tiny{}55.18} & {\tiny{}121} & {\tiny{}9.79} & {\tiny{}66}\tabularnewline
\cline{2-7} 
 & {\tiny{}Pashto} & {\tiny{}959} & {\tiny{}70.26} & {\tiny{}121} & {\tiny{}9.95} & {\tiny{}49}\tabularnewline
\cline{2-7} 
 & {\tiny{}Turkish} & {\tiny{}963} & {\tiny{}68.98} & {\tiny{}121} & {\tiny{}9.76} & {\tiny{}66}\tabularnewline
\cline{2-7} 
 & {\tiny{}Vietnamese} & {\tiny{}954} & {\tiny{}78.62} & {\tiny{}120} & {\tiny{}10.9} & {\tiny{}131}\tabularnewline
\cline{2-7} 
 & {\tiny{}Haitian} & {\tiny{}724} & {\tiny{}60.11} & {\tiny{}120} & {\tiny{}10.63} & {\tiny{}60}\tabularnewline
\cline{2-7} 
 & {\tiny{}Tamil } & {\tiny{}724} & {\tiny{}62.11} & {\tiny{}121} & {\tiny{}11.61} & {\tiny{}49}\tabularnewline
\cline{2-7} 
 & {\tiny{}Kurdish} & {\tiny{}502} & {\tiny{}37.69} & {\tiny{}120} & {\tiny{}10.21} & {\tiny{}64}\tabularnewline
\cline{2-7} 
 & {\tiny{}Tokpisin} & {\tiny{}482} & {\tiny{}35.32} & {\tiny{}120} & {\tiny{}9.88} & {\tiny{}55}\tabularnewline
\cline{2-7} 
 & {\tiny{}Georgian} & {\tiny{}490} & {\tiny{}45.35} & {\tiny{}120} & {\tiny{}12.30} & {\tiny{}35}\tabularnewline
\hline 
\multirow{2}{*}{{\tiny{}Target}} & {\tiny{}Assamese} & {\tiny{}720} & {\tiny{}54.35} & {\tiny{}120} & {\tiny{}9.58} & {\tiny{}66}\tabularnewline
\cline{2-7} 
 & {\tiny{}Swahili} & {\tiny{}491} & {\tiny{}40.0} & {\tiny{}120} & {\tiny{}10.58} & {\tiny{}56}\tabularnewline
\hline 
\end{tabular}{\tiny \par}
}
\caption{Details of the BABEL data used for experiments.}
\label{tab:data}
\end{table}

\section{Sequence to sequence model setup}
\label{sec:am-setup}
The baseline systems are built on
80-dimensional Mel-filter bank (fbank) features extracted from the speech samples using a sliding window of size 25 ms with 10ms stride. 
KALDI toolkit \cite{povey2011kaldi} is used to perform
the feature processing. 
The ``fbank'' features are then fed to a seq2seq model with the following configuration:

The Bi-RNN \cite{schuster1997bidirectional} models mentioned above uses an LSTM \cite{hochreiter1997long} cell followed by a projection layer (BLSTMP). 
In our experiments below, we use only a character-level seq2seq model
based on CTC and attention, as mentioned above.
Thus, in the following experiments, we will use character error rate (\% CER) as a suitable measure to analyze the model performance. 
All models are trained in ESPnet, end-to-end speech processing toolkit \cite{watanabe2018espnet}.




\section{Multilingual features}
\label{sec:MultFE}
Multilingual features are trained separately from seq2seq model
according to a setup from our previous RNN/DNN-HMM work~\cite{karafiat:SLT:2016:MultBLSTM}.
It allows us to easily combine traditional DNN techniques with the seq2seq model such as  GMM based
alignments for NN target estimation,
phoneme units and  frame-level randomization.
Multilingual features incorporate additional knowledge from non-target languages
into  features which could better guide the seq2seq model.

\subsection{Stacked Bottle-Neck feature extraction}
\label{sec:SBN}
The original idea of Stacked Bottle-Neck feature extraction is
described in~\cite{karafiat:IS:2014}. The scheme 
consists of two NN stages: 
The first one is 
reading short temporal context, its output is stacked, down-sampled,
and fed into the second NN reading longer temporal information.

The first stage bottle-neck NN input features are  24 log Mel filter
bank outputs concatenated with fundamental frequency
features.
Conversation-side based mean subtraction is  applied and 11
consecutive frames are  stacked. Hamming window followed by discrete
cosine transform (DCT) retaining 0$^{th}$ to 5$^{th}$ coefficients are  applied
on the time trajectory of each parameter resulting in 37$\times$6=222
coefficients at the first-stage NN input. 

In this work, the first-stage NN has 4 hidden layers with 1500
units in each except the bottle-neck (BN) one. The BN layer has 80
neurons. The neurons in the BN layer have linear activations as found
optimal in~\cite{vesely:asru:2011}. 21 consecutive frames from the first-stage NN are  stacked, down-sampled (each 5 frame is  taken) and fed into the second-stage  NN with an 
architecture similar to the first-stage NN, except of BN layer with only 30 neurons. Both neural networks were trained jointly as suggested in~\cite{vesely:asru:2011} in
CNTK toolkit~\cite{agarwal:TechRep:2014:CNTK} with block-softmax final
layer~\cite{vesely:slt:2012}. 
Context-independent phoneme states are used as the training targets
for the feature-extraction NN, otherwise the size of the final layer
would be prohibitive.  

Finally,  BN outputs from the second stage NN are used as features 
for further experiments and will be noted as ``Mult11-SBN".


\subsection{Results}

\begin{figure}[tb]
  \centering
  \centerline{\includegraphics[width=1.1\linewidth]{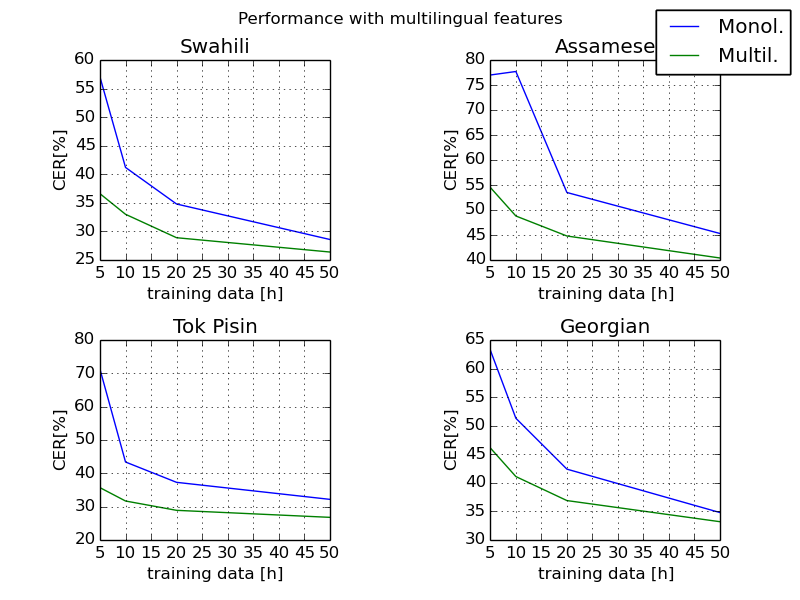}}
  \vspace{-2mm}
  \caption{Monolingual models trained on top of multilingual features.}
  \label{fig:MultFE}
\end{figure}

Figure~\ref{fig:MultFE} presents the performance curve of the seq2seq model with four ``train'' and ``target'' languages, as discussed in Section \ref{sec:data}, by changing the amount of training data.
It shows significant  performance drop of  baseline, ``fbank'' based, systems when the amount of training data is lowered.

On the other hand, the multilingual features present: 
1) significantly smaller performance degradation than baseline ``fbank'' features on small amounts of training data. 
2) consistent improvement on both train (seen) and target (unseen) languages where we only use train (seen) languages in feature extractor training
data. 3) significant improvement even on
the full training set, i.e., 1.6\%-5.0\% absolute (Table~\ref{tab:MultFE} summarizes the full training set results).

\begin{table}[tb]
\begin{tabular}{|l|c|c|c|c|}
\hline
\multirow{2}{*}{Features} 
            &  Swahili & Amharic & Tok Pisin & Georgian \\
            &  \%CER   & \%CER   & \%CER     & \%CER   \\   
\hline
FBANK       & 28.6           & 45.3            & 32.2 & 34.8\\ 
Mult11-SBN  & 26.4           & 40.4            & 26.8 & 33.2\\
\hline
\end{tabular}
\caption{Monolingual models trained on top of multilingual features.}
\label{tab:MultFE}
\end{table}

\section{Multilingual models}
\label{sec:MultAM}

Next, we focus on the training of multilingual seq2seq models.
As our models are character-based, the multilingual training dictionary is
created by concatenation of all train languages, and the system is trained
in same way as monolingual on concatenated  data.

\subsection{Direct decoding from multilingual NN}
As the multilingual net is trained to convert a sequence of input
features into sequence of output characters, any language from training
set or an unknown language with compatible set of 
characters can be directly decoded. Obviously, characters from
wrong language can be generated as the system needs to performs also
language identification (LID). Adding LID information as an 
additional feature,  similarly to~\cite{Kim_Seltzer:ICASSP:2018},
complicates the system. Therefore, we experimented with ``fine-tuning''
of the system into the target language by running a few  epochs only on
desired language data. This is in strengthening the target language
characters,  therefore it makes the system less prone 
to language- and character-set-mismatch  errors. 

The first two rows of table~\ref{tab:MultAM} present significant performance degradation from monolingual to multilingual seq2seq models caused by wrong decision of output characters in about 20\% of test utterances. 
However, no out-of-language characters are observed
after ``fine-tuning'' and 1.5\% and 4.7\% improvement over monolingual
baseline is reached.

\begin{table}[bh]
\begin{center}
\begin{tabular}{|l|c|c|}
\hline
\multirow{2}{*}{Model} 
            &  Tok Pisin & Georgian \\
            &  \%CER     & \%CER   \\   
\hline
Monolingual   & 32.2            & 34.8 \\
Multilingual  & 37.2            & 51.1 \\
Multilingual-fine tuned & {\bf 27.5}    & {\bf 33.3} \\
\hline 
\end{tabular}
\caption{Multilingual fine tuning of seq2seq model.}
\label{tab:MultAM}
\end{center}
\end{table}

\begin{figure}[tb]
  \centering
  \centerline{\includegraphics[width=1.\linewidth]{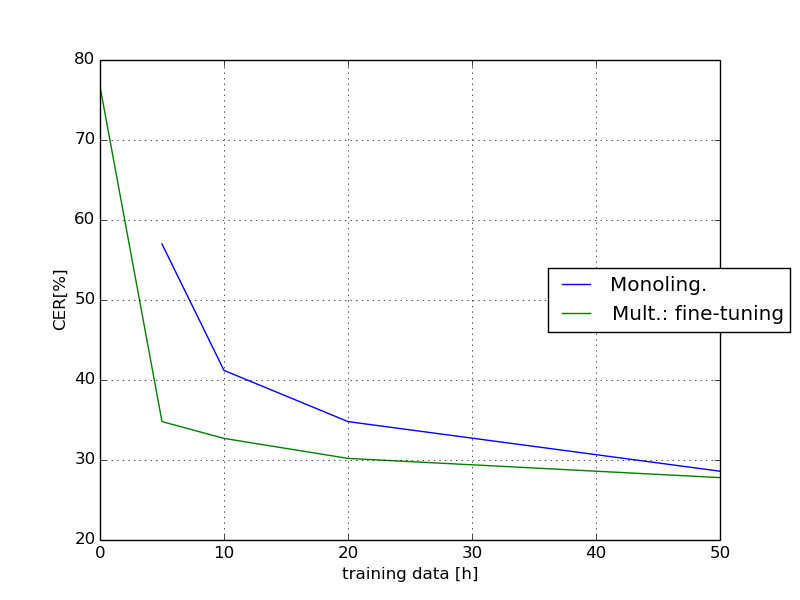}}
  \caption{Fine-tuning of multilingual NN on Swahili.}
  \label{fig:MultFT_swahili}
\end{figure}

As mentioned above, multilingual NN can be fine-tuned to the target language if character set is compatible with the training set. 
Figure~\ref{fig:MultFT_swahili} shows results on Swahili, which is not
part of the training set. Similarly to experiments with multilingual
features in Figure \ref{fig:MultFE}, the multilingual seq2seq systems are effective
especially on small amounts of data, but also beat baseline models on
full $\sim$50h language set.

\subsection{Language-Transfer learning}
\label{sec:LTL}

Language-Transfer learning is necessary if target language characters
differ from train set ones. The whole process can be described
in three steps: 1) randomly initialize  output layer parameters, 
 2) train only new parameters and freeze the remaining ones 3) ``fine-tune'' the whole NN.
Various experiments are conducted on level of output
parameters including output softmax (Out), the attention
(Att), and CTC parts.
Table~\ref{tab:MultLT} compares  all combinations and
clearly shows that retraining of output softmax only is giving
the best results. 

\begin{table}[bh]
\begin{tabular}{|l|c|c|c|c|}
\hline
Language    &  Swahili & Amharic & Tok Pisin & Georgian \\
 Transfer   &  \%CER   & \%CER   & \%CER     & \%CER   \\   
\hline
Monoling.   & 28.6           & 45.3            & 32.2 & 34.8\\ 
\hline 
Out                      & {\bf 27.4}     & {\bf 41.2}            & {\bf 27.7} & {\bf 33.6}\\
Att+Out                  & 27.5           & 41.2            & 28.3 & 34.2\\
CTC+Out                  & 27.6           & 41.2            & 27.9 & 33.7\\
Att+CTC+Out              & 28.0           & 42.1            & 27.6 & 34.1\\ 
\hline
\end{tabular}
\caption{Multilingual Language Transfer}
\label{tab:MultLT}
\end{table}



\begin{figure}[tb]
  \centering
  \centerline{\includegraphics[width=1.\linewidth]{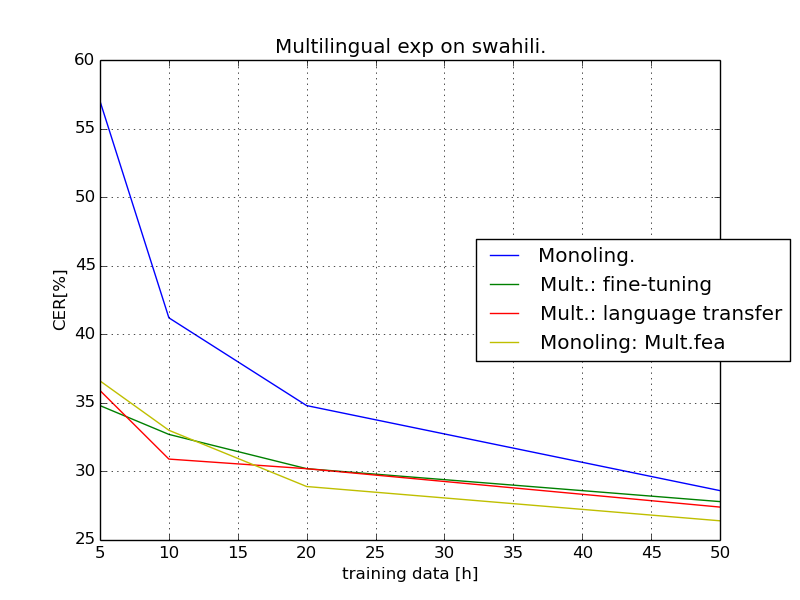}}
  \caption{Comparison of various multilingual approaches on Swahili.}
  \label{fig:Mult_swahili}
\end{figure}


Finally, we summarize the comparison of the use of multilingual features for the seq2seq model and language transfer learning of multilingual seq2seq model in Figure ~\ref{fig:Mult_swahili}.
Interestingly, on contrary to our previous observations on DNN-HMM systems \cite{karafiat:SLT:2016:MultBLSTM}, we found multilingual features superior to language transfer learning in seq2seq model case.

\section{Conclusions}
We have presented various multilingual approaches in seq2seq systems including multilingual features and multilingual models by leveraging our multilingual DNN-HMM expertise.
Unlike DNN-HMM systems \cite{karafiat:SLT:2016:MultBLSTM}, we obtain the opposite conclusion that multilingual features are more effective in seq2seq systems. 
It is probably due to efficient fusion of two complementary approaches: explicit GMM-HMM alignment incorporated in BN features and seq2seq models in the final system.
With this finding, we will further explore efficient combinations of the DNN-HMM and seq2seq systems as our future work.





\bibliographystyle{IEEEbib}
\bibliography{karaf,karthick}

\end{document}